\documentclass[twocolumn,preprintnumbers,amsmath,amssymb]{revtex4}

\usepackage{graphicx}% Include figure files
\usepackage{dcolumn}% Align table columns on decimal point

\usepackage{bm}% bold math
\usepackage[english]{babel}
\begin{document}

\preprint{FERMILAB-Pub-04/102-E}

\title{\boldmath New Measurement of the Top Quark Mass
in Lepton+Jets $t\overline{t}$ Events at D\O\ }
% Force line breaks with \\

% LIST_OF_AUTHORS_R1.TEX                 4/20/04            
%
\author{                                                                      
%% names begin here                                                           
V.M.~Abazov,$^{21}$                                                           
B.~Abbott,$^{54}$                                                             
A.~Abdesselam,$^{11}$                                                         
M.~Abolins,$^{47}$                                                            
V.~Abramov,$^{24}$                                                            
B.S.~Acharya,$^{17}$                                                          
D.L.~Adams,$^{52}$                                                            
M.~Adams,$^{34}$                                                              
S.N.~Ahmed,$^{20}$                                                            
G.D.~Alexeev,$^{21}$
% ...
\\                                                                            
\vskip 0.30cm                                                                 
\centerline{(D\O\ Collaboration)}                                             
\vskip 0.30cm                                                                 
}                                                                             
\address{                                                                     
\centerline{$^{1}$Universidad de Buenos Aires, Buenos Aires, Argentina}       
\centerline{$^{2}$LAFEX, Centro Brasileiro de Pesquisas F{\'\i}sicas,         
                  Rio de Janeiro, Brazil}                                     
\centerline{$^{3}$Universidade do Estado do Rio de Janeiro,                   
                  Rio de Janeiro, Brazil}                                     
\centerline{$^{4}$Institute of High Energy Physics, Beijing,                  
                  People's Republic of China}                                 
\centerline{$^{5}$Universidad de los Andes, Bogot\'{a}, Colombia}             
\centerline{$^{6}$Charles University, Center for Particle Physics,            
                  Prague, Czech Republic}                                     
\centerline{$^{7}$Institute of Physics, Academy of Sciences, Center           
                  for Particle Physics, Prague, Czech Republic}               
\centerline{$^{8}$Universidad San Francisco de Quito, Quito, Ecuador}         
\centerline{$^{9}$Laboratoire de Physique Subatomique et de Cosmologie,       
                  IN2P3-CNRS, Universite de Grenoble 1, Grenoble, France}     
\centerline{$^{10}$CPPM, IN2P3-CNRS, Universit\'e de la M\'editerran\'ee,     
                  Marseille, France}                                          
\centerline{$^{11}$Laboratoire de l'Acc\'el\'erateur Lin\'eaire,              
                  IN2P3-CNRS, Orsay, France}                                  
\centerline{$^{12}$LPNHE, Universit\'es Paris VI and VII, IN2P3-CNRS,         
                  Paris, France}                                              
\centerline{$^{13}$DAPNIA/Service de Physique des Particules, CEA, Saclay,    
                  France}                                                     
\centerline{$^{14}$Universit{\"a}t Freiburg, Physikalisches Institut,         
                  Freiburg, Germany}                                          
\centerline{$^{15}$Panjab University, Chandigarh, India}                      
\centerline{$^{16}$Delhi University, Delhi, India}                            
\centerline{$^{17}$Tata Institute of Fundamental Research, Mumbai, India}     
\centerline{$^{18}$CINVESTAV, Mexico City, Mexico}                            
\centerline{$^{19}$FOM-Institute NIKHEF and University of                     
                  Amsterdam/NIKHEF, Amsterdam, The Netherlands}               
\centerline{$^{20}$University of Nijmegen/NIKHEF, Nijmegen, The               
                  Netherlands}                                                
\centerline{$^{21}$Joint Institute for Nuclear Research, Dubna, Russia}       
\centerline{$^{22}$Institute for Theoretical and Experimental Physics,        
                   Moscow, Russia}                                            
\centerline{$^{23}$Moscow State University, Moscow, Russia}                   
\centerline{$^{24}$Institute for High Energy Physics, Protvino, Russia}       
\centerline{$^{25}$Lancaster University, Lancaster, United Kingdom}           
\centerline{$^{26}$Imperial College, London, United Kingdom}                  
\centerline{$^{27}$University of Arizona, Tucson, Arizona 85721}              
\centerline{$^{28}$Lawrence Berkeley National Laboratory and University of    
                  California, Berkeley, California 94720}                     
\centerline{$^{29}$California State University, Fresno, California 93740}     
\centerline{$^{30}$University of California, Irvine, California 92697}        
\centerline{$^{31}$University of California, Riverside, California 92521}     
\centerline{$^{32}$Florida State University, Tallahassee, Florida 32306}      
\centerline{$^{33}$Fermi National Accelerator Laboratory, Batavia,            
                   Illinois 60510}                                            
\centerline{$^{34}$University of Illinois at Chicago, Chicago,                
                   Illinois 60607}                                            
\centerline{$^{35}$Northern Illinois University, DeKalb, Illinois 60115}      
\centerline{$^{36}$Northwestern University, Evanston, Illinois 60208}         
\centerline{$^{37}$Indiana University, Bloomington, Indiana 47405}            
\centerline{$^{38}$University of Notre Dame, Notre Dame, Indiana 46556}       
\centerline{$^{39}$Iowa State University, Ames, Iowa 50011}                   
\centerline{$^{40}$University of Kansas, Lawrence, Kansas 66045}              
\centerline{$^{41}$Kansas State University, Manhattan, Kansas 66506}          
\centerline{$^{42}$Louisiana Tech University, Ruston, Louisiana 71272}        
\centerline{$^{43}$University of Maryland, College Park, Maryland 20742}      
\centerline{$^{44}$Boston University, Boston, Massachusetts 02215}            
\centerline{$^{45}$Northeastern University, Boston, Massachusetts 02115}      
\centerline{$^{46}$University of Michigan, Ann Arbor, Michigan 48109}         
\centerline{$^{47}$Michigan State University, East Lansing, Michigan 48824}   
\centerline{$^{48}$University of Nebraska, Lincoln, Nebraska 68588}           
\centerline{$^{49}$Columbia University, New York, New York 10027}             
\centerline{$^{50}$University of Rochester, Rochester, New York 14627}        
\centerline{$^{51}$State University of New York, Stony Brook,                 
                   New York 11794}                                            
\centerline{$^{52}$Brookhaven National Laboratory, Upton, New York 11973}     
\centerline{$^{53}$Langston University, Langston, Oklahoma 73050}             
\centerline{$^{54}$University of Oklahoma, Norman, Oklahoma 73019}            
\centerline{$^{55}$Brown University, Providence, Rhode Island 02912}          
\centerline{$^{56}$University of Texas, Arlington, Texas 76019}               
\centerline{$^{57}$Rice University, Houston, Texas 77005}                     
\centerline{$^{58}$University of Virginia, Charlottesville, Virginia 22901}   
\centerline{$^{59}$University of Washington, Seattle, Washington 98195}       
}                                                                             
%end                                                                          

\date{\today}% It is always \today, today,
             %  but any date may be explicitly specified

\begin{abstract}

We present a new measurement of the mass of the top quark 
using lepton + jets $t \bar t$ events collected by the D\O\ experiment
in Run I of the Fermilab Tevatron Collider. The mass is extracted through
a comparison of each event with a leading-order matrix element that
depends on the top quark mass. The result is $M_t=180.1 \pm 3.6 \mbox{ (stat)}
\pm 3.9 \mbox{ (sys)} \mbox{ GeV/$c^2$}$. Combining this improved
measurement with our previous value from dilepton channels yields the
new D\O\ result $M_t=179.0 \pm 3.5 \mbox{ (stat)} \pm 3.8 \mbox{ (sys)}
\mbox{ GeV/$c^2$}$.

\end{abstract}

% PACS, the Physics and Astronomy
                             % Classification Scheme.
%\keywords{Suggested keywords}%Use showkeys class option if keyword
                              %display desired
\maketitle

The observation of the top ($t$) quark  \cite{cdfPRD1,d0PRD2} was one of the major
confirmations  of the validity of the standard model (SM) of particle
interactions. Through radiative corrections of the SM,
the mass of the top quark ($M_t$), 
along with that of the $W$ boson ($M_W$) \cite{pdg3} , constrains the
 mass of the hypothesized Higgs
boson \cite{lepEWWG4}. $M_W$ is known to a precision of $< 0.1$\%,
while the uncertainty on $M_t$ is at the 3$\%$
level \cite{pdg3}. Improvements 
in both measurements are required to further limit the mass range of
the Higgs boson, and to check
the self-consistency of the SM. It is therefore important to
develop techniques for extracting a more precise value of $M_t$.

We report a new measurement of the mass of the top quark using
$t \bar t$ events containing an isolated lepton and four jets,
collected by the D\O\ experiment \cite{d0NIM6} in
Run I of the Fermilab Tevatron Collider. The data correspond to an
integrated luminosity of 125 pb$^{-1}$, and this
analysis is based on the same data sample used to extract 
$M_t$ in a previous publication \cite{massPRD5}.

As before, we assume that the top quark decays 100\% of the time
to a $W^+$ boson and a $b$ quark, which for a $t \bar t$ pair
implies $W^+ W^- b \bar b $ in the final state.
This analysis is based on decay channels containing a lepton 
(electron or muon from one $W \rightarrow l \nu_l$ decay )
and jets (from the evolution of the $b$ quarks and
the quarks from the other $W \rightarrow q \bar q'$ decay) in the final state \cite{massPRD5}.
After offline selections on lepton transverse energy ($E_T^{lep} > 20$ GeV) and
pseudorapidities ($|\eta_{\mu}| < 1.7$ for muons and $|\eta_{e}| < 2.0$ for
electrons), on jet transverse energies ($E_T > 15$ GeV) and pseudorapidities
($|\eta| < 2.0$), imbalance in transverse energy ($\not\!\! E_T > 20$ GeV),
and on $W$ boson decay products ($ E^{lep}_T +  \!\! \not\!\! E_T > 60$ GeV)
and pseudorapidity ($|\eta_W| < 2.0$),
and after applying several less-important criteria \cite{massPRD5}, the
original event sample consists of 91 events with one isolated lepton and four or
more jets. (Unlike the previous analysis, we do not distinguish between
events that have or lack a muon associated with one of the jets, signifying
the possible presence of a $b$-quark jet in the final state.) The new
analysis involves a comparison of these 91 events with a leading-order matrix
element for $t\overline{t}$ 
production and decay. To minimize the effect
of higher-order corrections, we restrict the study to events
containing exactly four jets, which reduces the sample to 71 events.

In the previous analysis, the four jets with highest $E_T$ were assumed
to represent the four quarks in the event. These, along with the lepton and the unobserved neutrino
were fitted to the kinematic
hypothesis $p \bar p \rightarrow t \bar t \rightarrow W W b \bar b$,
subject to the constraints of overall momentum-energy
conservation, the known mass of the $W$ boson, and the fact that
the unknown mass of the top quark was assumed to be identical for
the top and antitop quarks in the event. With twelve ways to permute
the jets, there were twelve possible fits (six when one of the 
jets was tagged as a $b$ jet), and the
solution with lowest $\chi^2$  was chosen as the best
hypothesis, thereby defining the 
fitted mass $m_{\text{fit}}$ for the event. The same procedure was
used to generate templates in variables of interest as a function
of input top quark mass. This was based on the {\sc HERWIG}
Monte Carlo (MC) program \cite{herwig7}, 
which was used to generate events that were
passed through full detector simulation and event reconstruction
\cite{d0geant8}.  Background events, consisting mainly of 
multijets (20$\%$) and $W$+jets (80$\%$) production, 
were processed in a similar manner. The
background from multijet production was based on studies of
multijet events in data \cite{massPRD5}, and the
background from $W$+jets events was based on events generated with
{\sc VECBOS} \cite{vecbos9}. A four-variable discriminant ($D$)
defined the probability that an event
represented signal as opposed to background. A probability density was
defined as a function of the discriminant $D$ and $m_{\text{fit}}$,
and a comparison of data and MC via a likelihood was used to
determine the most likely mass of the top quark. The resulting
measurement is $M_t=173.3 \pm 5.6 \mbox{ (stat)} 
\pm 5.5 \mbox{ (sys)} \mbox{ GeV/$c^2$}$.

The new method is similar to that suggested for $t \bar t$ dilepton decay
channels \cite{dgk10}, and used in a previous mass analyses of dilepton events
\cite{dilep11}. A similar approach has also been suggested for the
measurement of the mass of the $W$ boson at LEP \cite{berends11.5}.  
Given $N$ events, the top quark mass is estimated by maximizing the
likelihood:
\begin{equation}
L(\alpha)= e^{-N\int P_m(x,\alpha) {\textup d}x} \prod_{i=1}^N P_m(x_i,\alpha)
\end{equation}
where $x_i$ is a set of variables needed to specify the $i$th measured
event, $P_m$ is the probability density for observing that event, and $\alpha$
represents the parameters to be determined (in this case
$\alpha$ is the mass of the top quark).  Detector and reconstruction
effects are taken into account in two ways.  Geometric acceptance, 
trigger efficiencies, and event selection
enter through a multiplicative function $A(x)$ that is independent of
$\alpha$, and relates the observed probability density
$P_m(x,\alpha)$ to the production probability $P(x,\alpha)$:
$P_m(x,\alpha) = A(x) P(x,\alpha)$.
Energy resolution and merging and splitting
of jets are taken into account in a ``transfer'' 
function, $W(y,x)$, discussed below.
The production probability density can be written as a convolution of the
calculable cross section and $W(y,x)$:
\begin{equation}
P(x,\alpha) = \frac{1}{\sigma(\alpha)} \int {\textup d}\sigma(y,\alpha) {\textup d}q_1 {\textup d} q_2 f(q_1)
f(q_2) W(y,x)
\label{G_P(x)}
\end{equation}
where $W(y,x)$, our general transfer function, 
is the normalized probability density 
that the measured set of variables $x$ arise
from a set of partonic variables $y$, 
${\textup d}\sigma(y,\alpha)$ is the partonic
differential cross section, and $f(q_i)$ are parton distribution 
functions for the incoming partons with longitudinal momenta $q_i$.
Dividing by $\sigma(\alpha)$, the total cross section  
for the process, ensures that $P(x,\alpha)$ is properly
normalized.  The integral in Eq.(\ref{G_P(x)}) sums over all
possible parton states leading to what is observed in the detector.

For the $t \bar t$ production probability, the measured angles of 
the jets and of the charged lepton are assumed to be the angles of the
partons in the final state.
Given the detector resolutions the electron energy is assumed to be exact,
and the muon energy is described by its known resolution \cite{muon}.
Evaluation of Eq.(\ref{G_P(x)}) for the $e+$jets channel
involves two incident parton energies (we take these partons to be quarks,
and ignore the $\approx$10\%  contribution from gluon fusion), and six
objects in the final state.  The integrations over the essentially fifteen
sharp variables (three components of electron momentum, eight jet angles,
and four equations of energy-momentum conservation),
leave five integrals that must be performed to obtain the probability
that any event represents $t \bar t$ production for some specified
value of top quark mass $M_t$:
\begin{equation}
\label{ttbar_Px}
\begin{aligned}
P_{t \bar t}&=\frac{1}{12 \sigma_{t \bar t}} \int {\textup d}\rho_1
{\textup d}m_1^2 {\textup d}M_1^2
{\textup d}m_2^2 {\textup d}M_2^2 \\ \nonumber
              &\times \sum_{\text{perm},\nu}
 |{\cal M}_{t \bar t}|^2 \frac{f(q_1)f(q_2)}{|q_1||q_2|} \Phi_6
W_{\text{jets}}(E_{\text{part}},E_{\text{jet}})
\end{aligned}
\end{equation}
For $|{\cal M}_{t \bar t}|^2$, we use the leading-order matrix 
element \cite{sparke}, $f(q_1)$ and
$f(q_2)$ are  CTEQ4M parton distribution functions for the incident
quarks \cite{cteq}, $\Phi_6$ is the phase-space 
factor for the six-object final state, and
the sum is over all twelve permutations 
of the jets (the permutation of the jets from $W$ boson decay 
was performed by symmetrizing the matrix element), 
and the up-to-eight possible neutrino solutions.
Conservation of transverse momentum is used to calculate the
transverse momentum of the neutrino.
$W_{\text{jets}}(E_{\text{part}},E_{\text{jet}})$ is the part of $W(y,x)$
that refers to the mapping between parton-level energies
$E_{\text{part}}$ and energies measured in the detector $E_{\text{jet}}$.
Four of the
variables chosen for integration ($m_1$, $M_1$, $m_2$ and $M_2$), 
namely the masses of the $W$ bosons and of
the top quarks in the event, are economical in computing time, because the value of
$|{\cal M}_{t \bar t}|^2$ is essentially negligible 
except at the peaks of the four Breit-Wigner terms in the matrix element.
$\rho_1$ is the energy of one of the quarks in the hadronic decay of one of the
$W$ bosons.  The narrow-width approximation is used
to integrate over the top quark masses, and Gaussian 
adaptive quadrature \cite{gauss4.5} 
is used to perform the three remaining integrals.
$W_{\text{jets}}(E_{\text{part}},E_{\text{jet}})$ is the product of four
functions $F(E^i_{\text{part}},E^i_{\text{jet}})$, one for each jet, with
a functional form of the
sum of two Gaussians, with parameters having
linear dependence on $E^i_{\text{part}}$.
The parameters used for $b$ quarks are
different from those for the lighter quarks, 
and there are therefore twenty jet energy parameters in all.
About 15,000 simulated $t \bar t$ events (generated with masses between
140 and 200 GeV/$c^2$ in {\sc HERWIG}, and processed through detector simulation) 
are used to determine the above twenty parameters.
For a final state with a muon, $W_{\text{jets}}$
is expanded to include the 
muon momentum resolution, and an integration over muon momentum is
included in Eq.(\ref{ttbar_Px}).
\begin{figure}
\begin{center}
\includegraphics[width=\columnwidth]{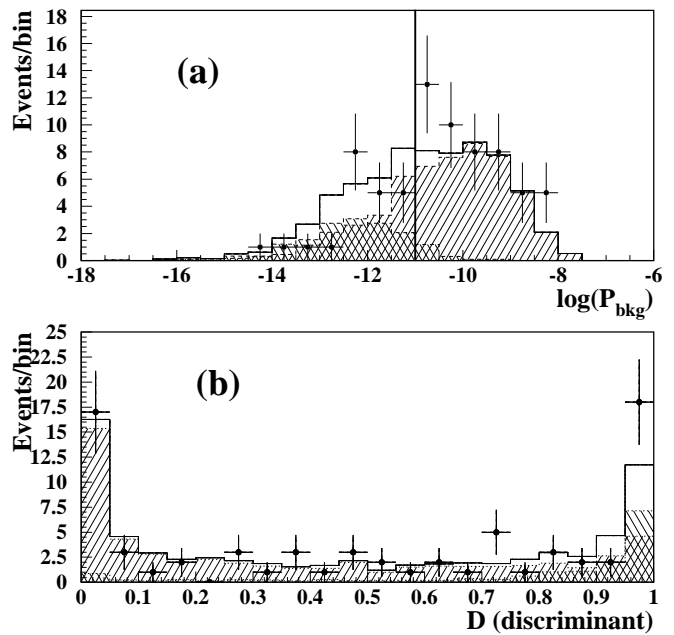}
\end{center}
\vskip -0.4cm
\caption[]{(a) Distribution in probability of events being background, and (b)
discriminant $P_{t\bar t}/(P_{t\bar t}+P_{\text{bkg}})$, calculated for the 71
$t\bar t$ candidates (data points). The data are compared with 
results expected for the sum (open histogram) from MC-simulated sources
of $t\overline{t}$
(left-hatched) and $W$+jets (right-hatched) events.
Only events with $P_{\text{bkg}}<10^{-11}$ (indicated by the vertical line)
are considered in the final analysis.}
\label{fig:probs}
\end{figure}

The $W$+4 jets matrix element from {\sc VECBOS}  
is used in Eq.(\ref{G_P(x)}) to calculate the background
probability $P_{\text{bkg}}$.  
The integration is performed over the energy of the four partons
leading to jets and the
$W$-boson mass.  The probability is summed over the twenty-four 
jet permutations and
two neutrino solutions.  The integration over parton energies is performed
using MC techniques, increasing the number of random points until 
the integral converges. (MC studies show that the 20\% background 
from multijet events is represented satisfactorily by that
for $W$+jets.)

After adding the probabilitites for the non-interfering $t\overline{t}$ and
$W$+4 jets channels,
the final likelihood as a function of $M_t$ is written as:
\begin{equation}
\label{tot_lik}
\begin{aligned}
-\ln & L(\alpha) = - \sum_{i=1}^N \ln[c_1 P_{t \bar t}(x_i, \alpha) +
c_2 P_{\text{bkg}}(x_i)] \\ \nonumber
         &+ N c_1 \int A(x) P_{t \bar t}(x, \alpha) {\textup d}x +
            N c_2 \int A(x) P_{\text{bkg}}(x) {\textup d}x
\end{aligned}
\end{equation}
The above integrals are calculated using MC methods, for which the
acceptance $A(x)$ is 1.0 or 0.0, depending on whether the event is
accepted or rejected by the analysis criteria.  
The best values of $\alpha$, representing the most
likely $M_t$, and the parameters $c_i$ are defined by minimizing
$-\ln L(\alpha)$.

Studies of samples of {\sc HERWIG} MC events used in the previous analysis
indicate that the new method should yield almost a factor of
two reduction in the statistical uncertainty on the extracted $M_t$.  These
studies also reveal that there is a systematic shift in $M_t$ that depends
on the amount of background in the data sample.  For high statistics, the shift
is about 2 GeV/$c^2$ when the background approaches 80$\%$ of total.  To
minimize this bias, a selection is introduced based on the probability
that an event represents background from $W+$jets.  Figure \ref{fig:probs}(a)
shows a comparison between the probability for a background interpretation
of a large sample of mixed MC events
(upper-most histogram) and the 71 $t \bar t$ candidates (data points).
The total number of MC events is normalized
to the 71 4-jet $t \bar t$ candidates.  The left-hatched (right-hatched)
histogram shows the contribution from $t \bar t$ ($W+4$ jets) MC events to the total.
Only the 22 events to the left of the vertical line are chosen for the final
analysis ($P_{\text{bkg}}<10^{-11}$).
The ratio of $t \bar t$ to $W+4$ jets events in the MC is normalized to
the $12/10$ ratio found for the data to the left of the vertical line, as described
below. (The selected value of $P_{\text{bkg}}<10^{-11}$ is based on MC studies carried
out before applying the method to data, and, for a top quark mass of 175
GeV/$c^2$, it retains 71$\%$ of the signal and 30$\%$ of the
background.)

A discriminant $D=P_{t\bar t}/(P_{t\bar t}+P_{\text{bkg}})$ was defined to
quantify the likelihood for an event to correspond to signal \cite{massPRD5}.
Since the signal probability depends on $M_t$, $D$ was calculated with the
signal probability taken at its most likely value.  Figure \ref{fig:probs}(b)
shows a comparison of the discriminant calculated for data (points with error
bars) and for MC events (open histogram), 
with the MC normalized as in
Fig. \ref{fig:probs}(a).
Since the discriminant depends directly on $M_t$, it was not
used to reject background and is shown simply to
illustrate the level of discrimination of signal from background.
%The discriminant was not used to reject background,
%because (unlike the background probability) its value depends directly on
%$M_t$, the parameter we are trying to determine. This discriminant is not
%used explicitly in this analysis, and is shown simply to illustrate the
%level of discrimination of signal from background.

\begin{figure}
\begin{center}
\includegraphics[width=\columnwidth]{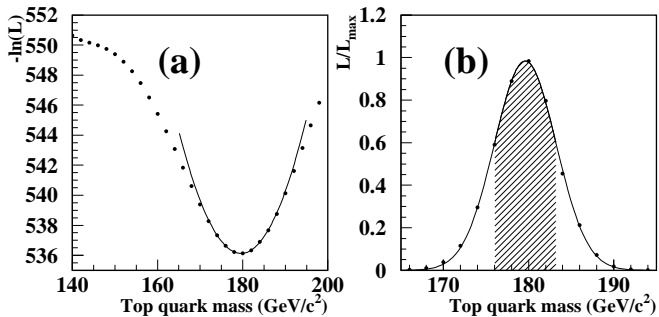}
\end{center}
\vskip -0.4cm
\caption[]{(a) Negative of the log of the likelihood as a function of the top quark mass. 
(b) The likelihood normalized to its maximum value in plot (a).
The curve is a Gaussian fit to the likelihood plot.
The hatched area corresponds to the 68.27$\%$ probability
interval.}
\label{fig:lik_topW}
\end{figure}

\begin{figure}
\begin{center}
\includegraphics[width=\columnwidth]{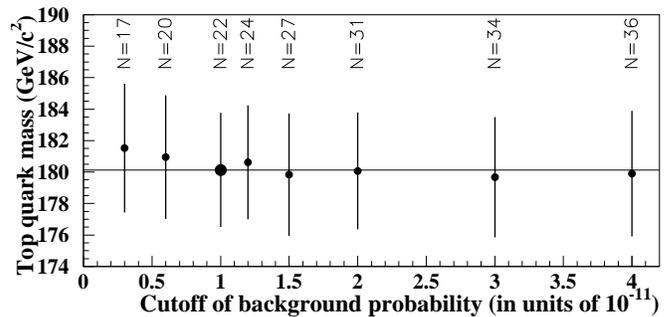}
\end{center}
\vskip -0.4cm
\caption[]{Mass of the top quark as a function of the cutoff in background
probability.  The number of remaining events is shown above each point.
The point with the larger dot is the value used in this analysis.}
\label{fig:back_stab}
\end{figure}

Figure \ref{fig:lik_topW}(a) shows the value of $-\ln L(\alpha)$ as a function
of $M_t$ for the 22 events that passed all selection criteria.
$-\ln L(\alpha)$ was minimized with respect to the parameters $c_i$ at each
mass point.  Figure \ref{fig:lik_topW}(b) shows the likelihood normalized to
its maximum value.  The Gaussian fit in the figure yields
$M_t = 179.6$ GeV/$c^2$, with an uncertainty $\delta M_t = 3.6$ GeV/$c^2$.
MC studies show that \cite{mass_note}: ($i$) $\delta M_t$ is compatible with
the uncertainties obtained in MC ensemble tests, and ($ii$) there is a
shift of -0.5 GeV/$c^2$ in the extracted mass.  After applying the 0.5 GeV/$c^2$
correction, our new value of the top quark mass is $M_t=180.1 \pm 3.6$ (stat) GeV/$c^2$.
As Fig. \ref{fig:probs}(a) indicates, the cutoff chosen in $P_{\text{bkg}}$ does not
reduce significantly the number of $t \bar t$ events, and therefore $M_t$
should be stable relative to variations in this cutoff.
Figure \ref{fig:back_stab} shows that a change in the cutoff in 
$P_{\text{bkg}}$ by more than
an order of magnitude changes the number of events used in the analysis by
more than a factor of two, but, as expected, does not have a significant impact
on $M_t$.

The total number of $t \bar t$ events in the 91-event sample is deduced to be
$(11\pm 3)/(0.71\times0.70\times0.87)=25 \pm 7$,  where 11 is the number
of extracted (using $c_1$ and $c_2$) events, and the corrections are for:
($i$) acceptance (0.71), ($ii$) events with more than 4 jets (0.70), 
and ($iii$)  $t \bar t$ 4-jets events that appear as background because a
leading-order matrix element does not represent them correctly (0.87). 
The number of $t \bar t$ events is consistent with 
that found in the previous analysis
\cite{massPRD5}.
The uncertainties in the quoted efficiencies are negligible compared
with the statistical uncertainty in the total number of 
$t\overline{t}$ events.

In the previous analysis \cite{massPRD5}, $\gamma+$jet events were used to
check the energy scale in the experiment relative to MC simulation.
This calibration had an uncertainty of $\delta E$=  0.025 $E$ + 0.5 GeV.
Consequently, we rescaled the energies 
of all jets in our sample by $\pm\delta E$,
redid the analysis, and the average of the two rescaled
results for $M_t$ ($\delta M_t =(3.0+3.5)/2 \approx 3.3$ GeV/$c^2$) 
is taken as the systematic error due to the
uncertainty in the jet energy scale (JES).  Additional contributions to 
the systematic uncertainty are listed in Table \ref{tb:sys}, and
details on the evaluation of these systematic 
errors and further description of the analysis technique 
can be found in Ref.\cite{mass_note}.

\begin{table}
\caption{Systematic uncertainties on the measurement of $M_t$.}
%\cite{massPRD5} }
\begin{center}
\renewcommand{\arraystretch}{1.4}
\setlength\tabcolsep{5pt}
\begin{tabular}{ll}
\hline
\noalign{\smallskip}
 Model for $t \bar t$               & 1.1 GeV/$c^2$ \\
 Model for backgound($W$+jets)      & 1.0 GeV/$c^2$ \\
 Noise and multiple interactions & 1.3 GeV/$c^2$ \\
 Jet energy scale                & 3.3 GeV/$c^2$ \\
 Parton distribution function    & 0.1 GeV/$c^2$ \\
 Acceptance correction           & 0.3 GeV/$c^2$ \\
 Bias correction                 & 0.5 GeV/$c^2$ \\
\hline
Total                            & 3.9 GeV/$c^2$ \\
\end{tabular}
\end{center}
\label{tb:sys}
\end{table}

Our method  provides substantial improvements in both the
statistical and systematic uncertainties.  This is due to two main differences
relative to the previous analysis: ($i$) each event now has an 
individual probability
as a function of the mass parameter, and therefore well-measured events
having a narrower likelihhod
contribute more to the extraction of the top quark mass than
those that are poorly measured, and ($ii$) all possible 
jet and neutrino combinations are included,
which guarantees that all 
signal
events contribute to the measurement.

In conclusion, we have presented a new measurement of the mass of the top
quark using a method that compares each individual event with the expected
differential cross section for $t \bar t$ production and decay. We obtain a
significant improvement in statistical uncertainty over the previous
measurement \cite{massPRD5} that is equivalent to a factor of 2.4 more data.

From the differences in the two analyses, and from statistical
fluctuations arising from using a subsample of the
original data, we expect the difference between the original and the
new mass measurement to be on the order of 4 GeV/$c^2$. 
Thus, the two results
differ by less than two standard deviations.
The current analysis is also less sensitive to the 
calibration of the JES, and leads to an
improved systematic uncertainty.  The new result is:
\begin{equation}
M_t=180.1 \pm 3.6 \mbox{ (stat)} \pm 3.9 \mbox{ (sys)} \mbox{ GeV/$c^2$}.   \nonumber
\end{equation}
Combining the two uncertainties in quadrature, we obtain $M_t=180.1 \pm 5.3$
GeV/$c^2$, which has an uncertainty comparable to all the previous
measurements of D\O\ and CDF \cite{pdg3} combined.

Using the procedure described in Ref.\cite{comb12},
the new measurement can be combined with that
obtained using the dilepton sample collected
at D\O\ during Run I \cite{dilep11}, yielding 
the new D\O\ value for the mass of the top quark:
\begin{equation}
M_t=179.0 \pm 3.5 \mbox{ (stat)} \pm 3.8 \mbox{ (sys)} \mbox{ GeV/$c^2$}   \nonumber
\end{equation}
This is the most accurate measurement of the top quark mass
in any single experiment. The impact of the new D\O\ top-quark mass measurement 
on the world average top-quark mass as well as on 
Higgs and supersymmetry constraints is a subject of a 
separate recent publication \cite{nature}.

%\section*{Aknowledgements}
% acknowledgement_paragraph_r1.tex            4/20/04
%
We thank the staffs at Fermilab and collaborating institutions, 
and acknowledge support from the 
Department of Energy and National Science Foundation (USA),  
Commissariat  \` a l'Energie Atomique and 
CNRS/Institut National de Physique Nucl\'eaire et 
de Physique des Particules (France), 
Ministry of Education and Science, Agency for Atomic 
   Energy and RF President Grants Program (Russia),
CAPES, CNPq, FAPERJ, FAPESP and FUNDUNESP (Brazil),
Departments of Atomic Energy and Science and Technology (India),
Colciencias (Colombia),
CONACyT (Mexico),
Ministry of Education and KOSEF (Korea),
CONICET and UBACyT (Argentina),
The Foundation for Fundamental Research on Matter (The Netherlands),
PPARC (United Kingdom),
Ministry of Education (Czech Republic),
A.P.~Sloan Foundation,
and the Research Corporation.

\end{document}